\documentclass[runningheads]{llncs}
\usepackage{graphicx, tabularx, longtable}
\usepackage[table,xcdraw]{xcolor}
\newcolumntype{Y}{>{\centering\arraybackslash}X}
\usepackage[urlcolor=blue]{hyperref}

\usepackage{cite}
\usepackage{amsmath}

\begin{document}
\title{PAT-CNN: Automatic Segmentation and Quantification of Pericardial Adipose Tissue from T2-Weighted Cardiac Magnetic Resonance Images}

\titlerunning{PAT-CNN}

% uncomment after anonymisation can be lifted
% \author{anonymous}
\author{Zhuoyu Li\inst{1} \and Camille Petri\inst{2, 3} \and James Howard\inst{3} \and Graham Cole\inst{3}  \and Marta Varela\inst{3}*}

% uncomment after anonymisation can be lifted
\authorrunning{Z. Li et al.}
% \authorrunning{anonymous}

% uncomment after anonymisation can be lifted
% \institute{anonymous}
\institute{
Department of Metabolism, Digestion \&Reproduction, Imperial College London, UK \and
Department of Computing, Imperial College London, UK \and
National Heart \& Lung Institute, Imperial College London, UK \\
\email{*marta.mvarela@imperial.ac.uk}}


%
\maketitle              % typeset the header of the contribution
\begin{abstract}

% \textbf{Background:} Increased pericardial adipose tissue (PAT) is a known risk factor for cardiovascular disease (CVD). Although cardiac magnetic resonance (CMR) images are often used in patients at risk of CVD, there are currently no tools to automatically identify and quantify PAT from CMR. The aim of this study was to create a neural network to segment PAT from T2-weighted CMR and explore the correlations between PAT volumes (PATV) with CVD outcomes and mortality.

\textbf{Background:} Increased pericardial adipose tissue (PAT) is associated with many types of cardiovascular disease (CVD). Although cardiac magnetic resonance images (CMRI) are often acquired in patients with CVD, there are currently no tools to automatically identify and quantify PAT from CMRI. The aim of this study was to create a neural network to segment PAT from T2-weighted CMRI and explore the correlations between PAT volumes (PATV) and CVD outcomes and mortality.

% \textbf{Methods:} We trained and tested a deep learning model, called PAT-CNN, to segment PAT in T2-weighted cardiac MR images, using 70 images using semi-automatic PAT labels and data augmentation. Using the segmentations from PAT-CNN, we automatically calculated PATV on images from 391 patients. We used PATV to predict CVD diagnosis and 1-year mortality post-imaging.

\textbf{Methods:} We trained and tested a deep learning model, PAT-CNN, to segment PAT on T2-weighted cardiac MR images. Using the segmentations from PAT-CNN, we automatically calculated PATV on images from 391 patients. We analysed correlations between PATV and CVD diagnosis and 1-year mortality post-imaging.

% \textbf{Results:} PAT-CNN was able to accurately segment PAT with Dice score/ Hausdorff distances on 10 test CMR images of $0.74 \pm 0.03/27.1 \pm 10.9~mm$. This values compare well to $0.76 \pm 0.06/21.2 \pm 10.3~mm$ inter-observer agreement. Pearson correlation coefficients and regression models showed that, independently of age and BMI, PATV is significantly positively correlated with a diagnosis of CVD and the risk of death in one year post scan (p-value \textless 0.01).

\textbf{Results:} PAT-CNN was able to accurately segment PAT with Dice score/ Hausdorff distances of $0.74 \pm 0.03/27.1 \pm 10.9~mm$, similar to the values obtained when comparing the segmentations of two independent human observers ($0.76 \pm 0.06/21.2 \pm 10.3~mm$). Regression models showed that, independently of sex and body-mass index, PATV is significantly positively correlated with a diagnosis of CVD and with 1-year all cause mortality (p-value \textless 0.01).

\textbf{Conclusions:} PAT-CNN can segment PAT from T2-weighted CMR images automatically and accurately. Increased PATV as measured automatically from CMRI is significantly associated with the presence of CVD and can independently predict 1-year mortality.

\keywords{Cardiac MRI \and Segmentation \and Pericardial Fat Volume \and Cardiovascular Disease Risk Factors \and Convolutional Neural Network}
\end{abstract}

\setlength\belowcaptionskip{-3.5ex}

\section{Introduction}
\label{label:introduction}
Large volumes of pericardial adipose tissue, PAT, and endocardial adipose tissue, EAT, predispose to a number of cardiovascular conditions \cite{2020Adipokines} such as atrial fibrillation \cite{Fitzgibbons2014EpicardialAssociations}. This is due to the high metabolic activity of adipose tissue which can stimulate remodelling in cardiac muscle, leading to fibrosis, chamber enlargement and inflammation, among others. PAT can also regulate cardiac tissue innervation \cite{Lavie2017THEExercise} likely pro-arrhythmically. Obese patients usually have large amounts of PAT, but correlations between cardiac fat volume and body mass index (BMI) are relatively weak \cite{Fitzgibbons2014EpicardialAssociations, Henningsson2020QuantificationMRI}.

% Obesity increases the overall risk of cardiovascular disease (CVD) and mortality. However, obesity cannot be simply represented by body mass index (BMI), as body fat distribution is also important. Pericardial adipose tissue (PAT) refers to the fat deposited around the myocardium and pericardium, which has been found to be strongly correlated to pathogenesis of CVDs \cite{2020Adipokines}, because of its
% unique properties and its proximity to cardiac structures \cite{Iozzo2011}. ADD MORE DETAILS

% \textbf{Cardiac MRI}
% Through medical imaging techniques, we can measure the thickness, volume, density and other characteristics of PAT \cite{verhaert2010}, which have been found to be clinically relevant [ref]. Although fat is typically easier to identify and segment from computed tomography (CT) images, there is a clinical need to segment it in CMR images, which are increasingly used to diagnose and manage CVD due to MRI's flexible contrast.

% [In the absence of dedicated fat suppression techniques, fat is typically bright in most MR images due to its comparatively low T1 and high T2. -> ADD ELSEWHERE]

So far, PAT has been predominantly segmented and quantified using computed tomography (CT) \cite{Bencevic2022}. In these images, fat's low Hounsfield number allows for comparatively straightforward identification and segmentation using simple intensity threshold-based methods such as region growing and active contours \cite{Rodrigues2016ATomography} or more complicated approaches like neural networks \cite{Commandeur2018DeepCT}. Fat is in general more difficult to segment in Cardiac Magnetic Resonance Images (CMRI), where its intensity is much more variable and similar to other tissues'. Quantification of PAT volume (PATV) from MRI has so far only been reported using time-consuming manual segmentation of PAT \cite{Davidovich2013ImagingFat} or dedicated CMRI sequences \cite{Ding2016} such as Dixon imaging \cite{Henningsson2020QuantificationMRI, Kellman2009MultiechoMyocardium}. This prevents widespread quantification of PATV from more common types of CMRI, despite CMRI's increasing prominence in the diagnosis and assessment of cardiovascular disease (CVD). A tool to rapidly and accurately quantify PAT from CMRI would allow rapid quantification of PATV in the clinic, to aid CVD risk assessment.
%Convolutional neural networks (CNNs) have shown great ability to segment MR images, enabling easy quantification of several imaging biomarkers which would otherwise require manual image analysis too time consuming for clinical applications.

\textbf{MRI segmentation}
Convolutional neural networks (CNNs) have shown great ability to segment MR images, enabling easy quantification of several imaging biomarkers whose analysis would otherwise be too time consuming. They consistently outperform non-learned techniques such as thresholding \cite{2014Norouzi}, region growing \cite{2005Mancas} and probability graph models \cite{2013Grosgeorge}. Once trained, CNNs can perform image analysis in real time in the clinic. To date, and to the best of our knowledge, there are no CNN-based tools for the segmentation of PAT from T2-weighted CMR images.

% However, manual segmentation of PAT from CMR images is difficult and time-consuming.
% Automatic segmentation techniques such as thresholding \cite{2014Norouzi}, region growing \cite{2005Mancas} and probability graph model \cite{2013Grosgeorge} have disadvantages of mass computing, slow speed, and low accuracy. The most used deep learning models for MRI segmentation are based on convolutional neural networks (CNN), fully convolutional networks (FCN) or UNet \cite{2020A}. Nevertheless, there are currently no tools to enable automatic PAT segmentation from CMRI.

\textbf{Aims}
In this project, we aim to estimate PATV from clinical CMRI. We focus on axial T2-weighted rapid gradient-echo scans which are routinely acquired in the clinic and in which CNNs have been used to accurately segment several cardiac structures \cite{Howard2021}. We initially create a non-learned semi-automatic image processing protocol to segment PAT and generate training data for the proposed CNN. We then train and test a dedicated CNN to segment PAT in these images. Finally, using data from 391 patients, we investigate correlations between: a) PATV and the presence of CVD and b) PATV and 1-year all-cause mortality. 

% \textbf{Aims}
% Because of the strong relationship between PAT and CVD risks, as well as the difficulties of PAT segmentation in CMRI, the purpose of this study was to create a convolutional neural network (CNN) that can automatically segment PAT from T2-weighted MRI scans and use it to explore correlations between PAT volumes (PATV) and: the presence of CVD at imaging and 1-year post-imaging mortality.

\section{Methods}
\label{label:methods}
\textit{\textbf{Data}}
We uses fully anonymized T2-prepared spoiled GRE axial stacks acquired at 1.5T (TE: 1.56 ms, TR: 3.1 ms, recon voxel size: 1.4 x 1.4 x 6.0 $mm^3$) in 391 patients with known or suspected CVD, under ethical approval. The number of slices in each volumetric scan ranged from 23 to 55.  

\begin{figure}[ht!]
\centering
  \includegraphics[width=0.9\textwidth]{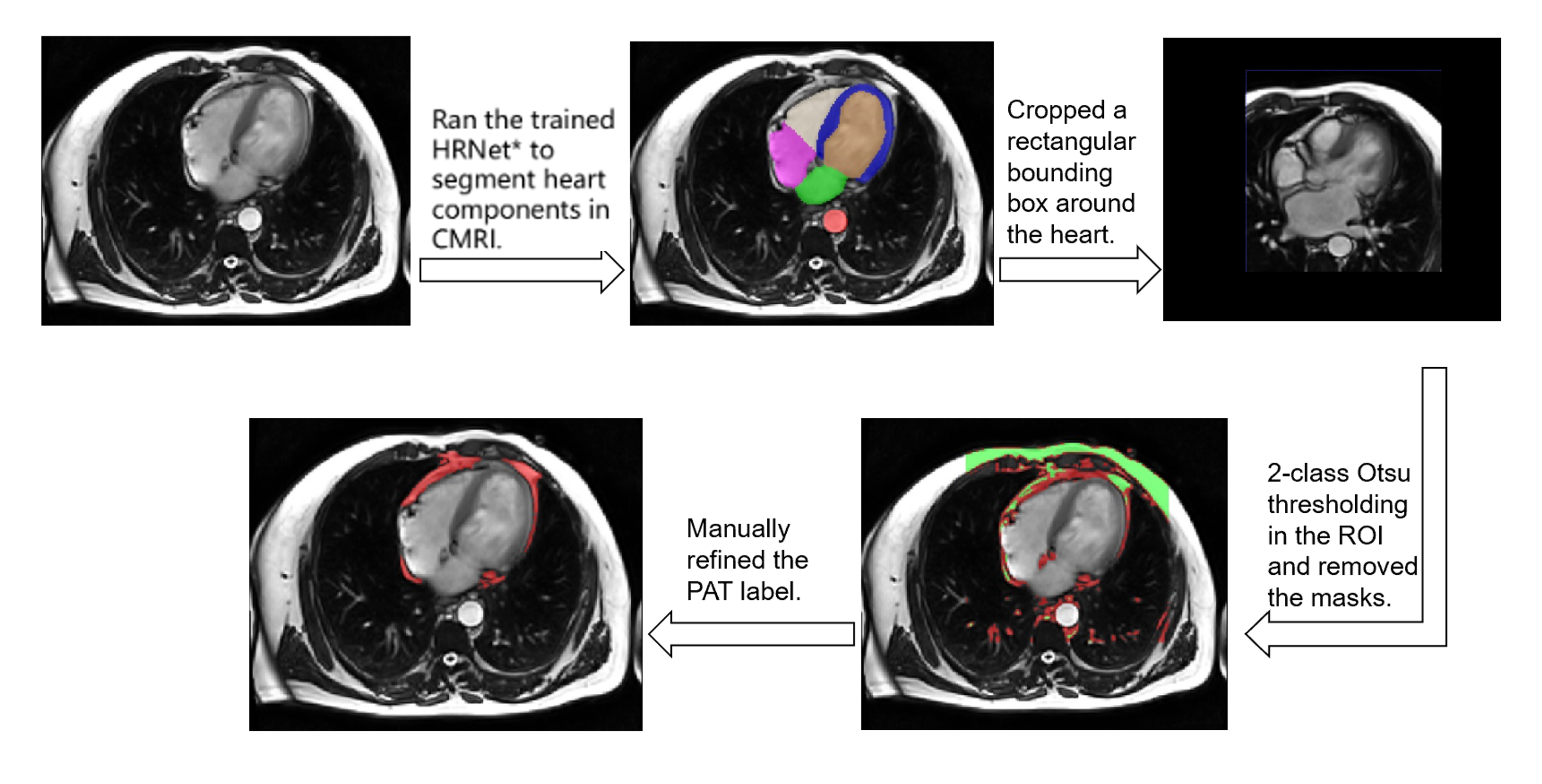}
  \caption{Semi-automatic pipeline for ground truth PAT segmentation. ROI: Region of Interest}
  \label{fig1}
\end{figure}

\textit{\textbf{Preprocessing}}
\label{label:methods:preprocessing}
Ground truth PAT segmentation was semi-automatically performed in 70 training images (Fig \ref{fig1}). As the first step, we identified the cardiac chambers and left ventricular (LV) myocardium using HRNet, an existing CNN trained in T2-weighted images with a similar contrast \cite{Howard2021}. Using these segmentations, we created a bounding box around the heart. We then performed a 2-class Otsu segmentation \cite{4310076} on the cropped images. In most cases, one of the classes roughly coincided with subcutaneous fat, allowing the rapid segmentation of PAT after some further manual corrections. These semi-automatic PAT segmentations were reviewed by a second independent observer. 

% As a final step, we resampled the images to the standardized size of 256 (width) * 256 (height) * 32 (slices) to be used as inputs to the network.

\begin{figure}[ht!]
\centering
  \includegraphics[width=0.9\textwidth]{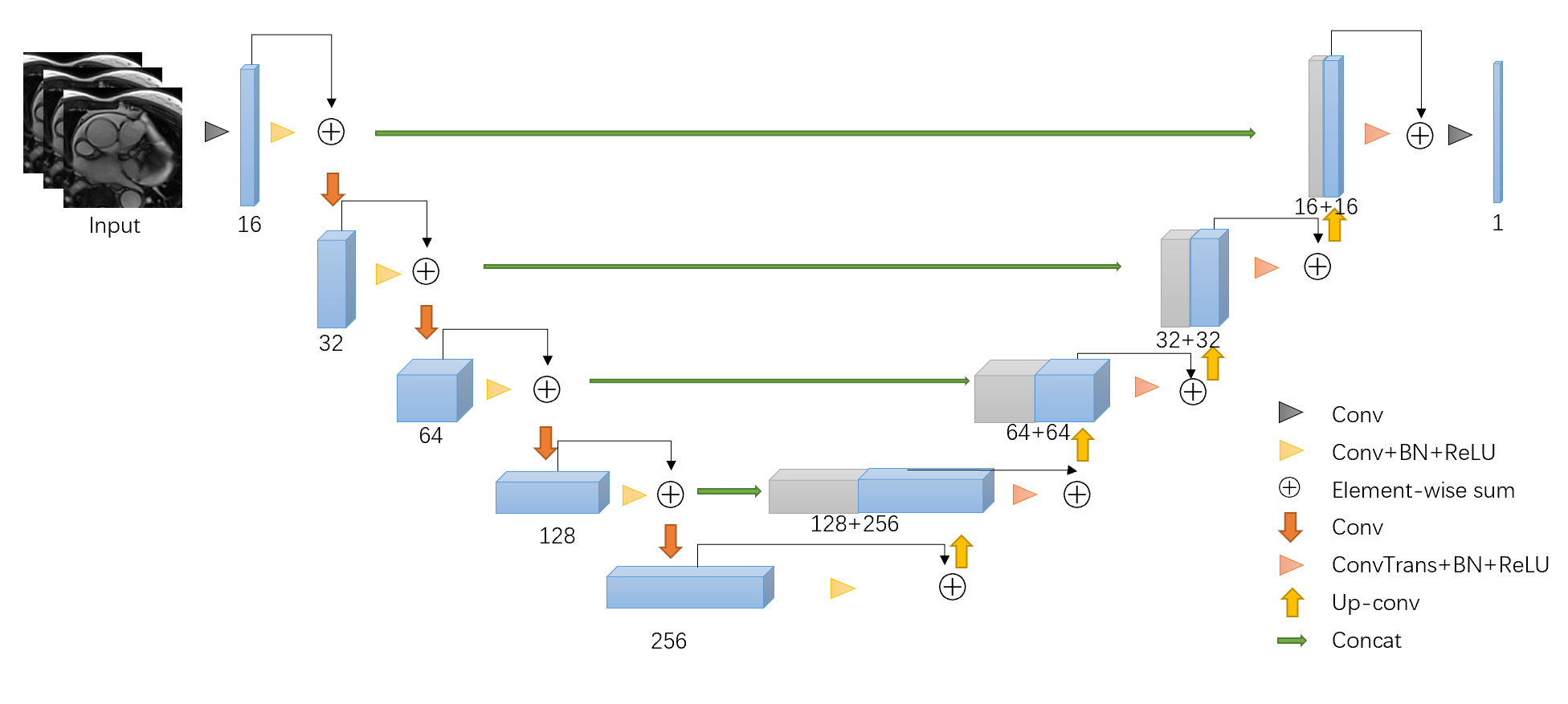}
  \caption{Structure diagram of PAT-CNN. The kernel sizes of convolutions and transpose convolutions are 3 x 3 x 3. The numbers represent filters numbers in each layer. The triangle arrows represents the residual units.}
  \label{fig2}
\end{figure}

\textit{\textbf{Model: Architecture}}
\label{label:methods:model}
To allow fully automatic PAT labelling, we trained PAT-CNN, a 3D Res-UNet CNN \cite{Kerfoot2019Left-VentricleU-Net}. The network (shown in Fig \ref{fig2}) included a contracting path with four convolutional layers, an expanding path with four upsampling layers and concatenate connections. A residual unit was added to each layer. In residual units, batch normalization and parametric rectified linear unit (PReLU) were implemented to improve the feature selection results. The output layer used a sigmoid function as the activation function and was trained to classify each pixel as: PAT (label 1) or background/other tissues (label 0).

\textit{\textbf{Model: Dataset Setup}}
70 scans with manual PAT labels were randomly divided into a training+validation dataset (60 images) and a testing dataset (10 images). A second independent observer independently segmented PAT in the test dataset to assess inter-observer segmentation metrics.

\textit{\textbf{Model: Data Augmentation}}
To enhance PAT-CNN's invariance and generalization, as well as expand the size of the dataset, we applied data augmentation during training using the image analysis library SimpleITK \cite{yaniv2018}. We randomly applied rotation, crop, flip, blur and Rayleigh noise addition to each input image of the network in each epoch. The probability of each transformation was 0.1.

\textit{\textbf{Model: Loss Function}}
The loss function combined cross-entropy loss and Dice loss terms \cite{Yeung2022UnifiedSegmentation}. The Dice loss term help improves PAT-CNN's accuracy in the presence of the large imbalance between the volumes of PAT and voxels labelled as background/other organs. The cross-entropy loss term minimises the likely instability of the gradients of a pure Dice-score-based loss function. The total loss function can be represented as below, where $x_{n}$ is the ground truth label of the $n^{th}$ sample, $y_{n}$ is its prediction, N is the total number of samples (and the small term $\epsilon$ in the Dice coefficient is used to avoid the denominator becoming 0):

\begin{equation}
    Loss(X, Y) = 1 - \frac{2 \sum_{n=1}^{N} x_{n}y_{n} + \epsilon}{\sum_{n=1}^{N} x_{n} + \sum_{n=1}^{N} y_{n} + \epsilon} - \frac{1}{N} \sum_{n=1}^{N} x_{n}log(y_{n})
\end{equation}

\textit{\textbf{Model: Hyperparameters and Performance Assessment}}
Through our experimental results, the optimal batch size was 8. We used Adam \cite{2014Adam} and set the learning rate to $10^{-4}$, training for 60 epochs. This combination performed best on the validation dataset.
We used the Dice Score and Hausdorff Distance to evaluate PAT-CNNs performance on the test dataset. Using these metrics, we additionally compared PAT-CNN to the vanilla U-Net \cite{Ronneberger2015} commonly using for segmenting medical images.

\textit{\textbf{Statistical Analysis: Data}}
We used PAT-CNN to estimate PATV (in $cm^3$) in all 391 volumetric T2-weighted axial CMR images. From the anonymized imaging metadata and radiology reports, we extracted  the following demographic information ($mean \pm std$): age ($55 \pm 18$ years), sex ($42\%$ female), BMI ($27.7 \pm 5.9~kg/m^2$) and 1-year post-scan mortality (`Deceased', $8.2\%$). We also compiled diagnostic information, which was synthesised into 73 binary clinical diagnostic labels (presence/absence of: dilated cardiomyopathy, left/right ventricular failure, left/right ventricular hypertrophy, myocardial infarction, ischaemia, etc). The sum of these diagnostic labels was compiled into a single variable `CVD Diagnosis' ($2.9 \pm 1.9$). 
%The dataset only records their survival after one year of their scans as binary data, so we did not have data for Mortality analysis.

\textit{\textbf{Statistical Analysis: Methods}} We investigated the relationship between the regressor variables: PATV, sex, age and BMI and the dependent variables: `CVD diagnosis', `Deceased' and PATV, using univariate Pearson/phi correlation analysis. Those regressors that were significantly correlated ($\alpha$ = 0.01) with the independent variables were then used as inputs to a multivariate analysis using ordinary least squares (for the `CVD Diagnosis' and `PATV' dependent variables) and a logistic regression model (for the dichotomous variable `Deceased'). We note that the clinical dataset used did not include time-to-mortality information and survival analyses were therefore not performed.

\section{Results}
\begin{figure}[ht!]
\centering
  \includegraphics[width=0.8\textwidth]{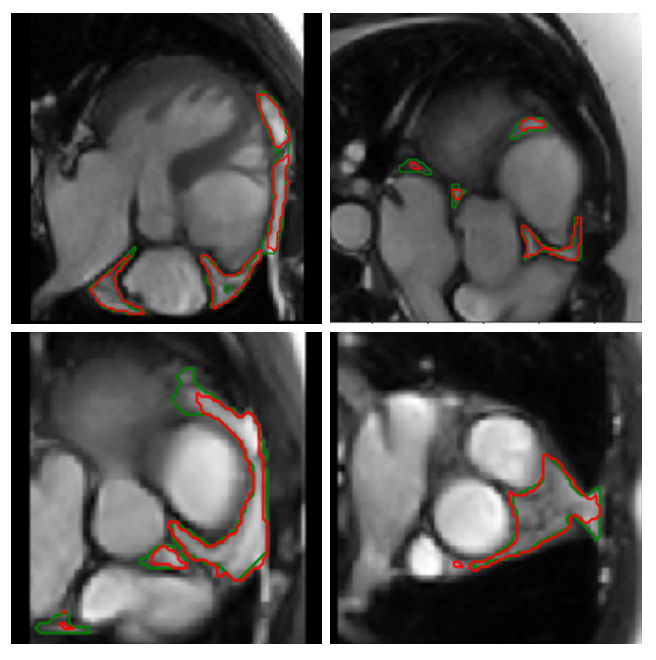}
  \caption{Representative examples of T2-weighted slices from four different patients overlaid with the contours for the ground truth (red) and PAT-CNN (green) PAT labels.}
  \label{fig3}
\end{figure}

\textit{\textbf{PAT-CNN Performance}}
On the test dataset, PAT was segmented with an average Dice score/Hausdorff distance of $0.74 \pm 0.03/27.1 \pm 10.9~mm$. The model coped well with variations in PAT shape, size and location (Fig \ref{fig3}) and the presence of pathological changes. PAT-CNN performed better than U-Net and its results compared well with the agreement between two human operators on the same dataset (Table \ref{tab:results1}). 

% \begin{table}[ht!]
% \centering
% \setlength{\abovecaptionskip}{0.15cm}
% \begin{tabularx}{\textwidth}{ |Y|Y|Y|Y| }
% \hline
%                     & \multicolumn{1}{c|}{\cellcolor[HTML]{ECF4FF}\textbf{Observer 1-PAT-CNN}} &  \multicolumn{1}{c|}{\cellcolor[HTML]{ECF4FF}\textbf{Observer 1-U-Net}} &  \multicolumn{1}{c|}{\cellcolor[HTML]{ECF4FF}\textbf{Observer 1-Observer 2}} \\ \hline
% \cellcolor[HTML]{ECF4FF}\textbf{Dice Score}              & \cellcolor[HTML]{FFFFFF}0.74 \pm 0.03                     & 0.71 \pm 0.03                     & 0.76 \pm 0.06                                                  \\ \hline
% \cellcolor[HTML]{ECF4FF}\textbf{Hausdorff Dist (mm)} & 27.1 \pm 10.9         & 32.8 \pm 11.2                    & 21.2 \pm 10.3                                                  \\ \hline
% \end{tabularx}
% \caption{PAT-CNN performance on the 10 images compared to the segmentation results between two human operators.}
% \label{tab:results1}
% \end{table}

% \usepackage{longtable}
% Note: It may be necessary to compile the document several times to get a multi-page table to line up properly

\begin{table}[ht!]
\centering
\setlength{\abovecaptionskip}{0.05cm}
\renewcommand{\arraystretch}{1}
\resizebox{\columnwidth}{!}{
\begin{tabular}{|c|c|c|c|}
\hline
                                                     & \cellcolor[HTML]{ECF4FF}\textbf{PAT-CNN} & \cellcolor[HTML]{ECF4FF}\textbf{U-Net} & \cellcolor[HTML]{ECF4FF}\textbf{Observer 2} \\ \hline
\cellcolor[HTML]{ECF4FF}\textbf{Dice Score}          & $0.74 \pm 0.03$                   & $0.71 \pm 0.03$                & $0.76 \pm 0.06$                    \\ \hline
\cellcolor[HTML]{ECF4FF}\textbf{Hausdorff Distance (mm)} & $27.1 \pm 10.9$         & $32.8 \pm 11.2$                    & $21.2 \pm 10.3$                         \\ \hline
\end{tabular}
}
\caption{Performance of PAT-CNN, U-Net and a second human observer on the test images. (The segmentations from human Observer 1 were treated as the ground truth for this analysis.)}
\label{tab:results1}
\end{table}

\textit{\textbf{Statistical Analysis}}
Across the 391 analysed patients, PATV had a mean of $139.60~cm^{3}$ and a standard deviation of $80.24~cm^{3}$, in excellent agreement with PAT quantification from CT \cite{Commandeur2018DeepCT}. PATV is significantly correlated with patients’ age ($\rho_{Pearson} = 0.38$), sex and BMI ($\rho_{Pearson} = 0.34$), but in a multivariate analysis, only sex and BMI remain significant (Table \ref{tab:results2}). PATV and age were the only regressor variables found to be significantly correlated to both CVD diagnostic labels and 1-year mortality post-imaging in both univariate and multivariate analyses (Table \ref{tab:results2}). This is in contrast to BMI which was not an independent predictor of either mortality or CVD diagnosis in univariate analysis (Table \ref{tab:results2}). 

% \begin{table}[ht!]
% \centering
% \renewcommand{\arraystretch}{1.2}
% \setlength{\abovecaptionskip}{0.3cm}
%  \resizebox{\textwidth}{!}{
% \begin{tabularx}{\textwidth}{Y|Y|Y|Y}
% \hline
% \textbf{Variables} & \textbf{PATV ($cm^{3}$)}  & \textbf{CVD diagnosis}   & \textbf{Deceased}            \\ \hline
% Sex  &  -50.2 (\textless 0.001)  &  -0.30 (0.10)   &  /    \\ \hline
% Age & 1.70 (\textless 0.001) & 0.02 (\textless 0.001) & 0.0273 (0.04)  \\ \hline
% BMI (kg/$m^{2}$) & 3.92 (\textless 0.001) & / & /    \\ \hline
% PATV ($cm^{3}$) & / & 0.01 (\textless 0.001) & 0.01 (\textless 0.001)  \\ \hline
% \end{tabularx}}
% \caption{Results from multivariate regression analyses (n = 391). Entries are regression coefficients (p-values from t-test at $\alpha = 0.05$) PATV: pericardial adipose tissue volume.}
% \label{tab:results2}
% \end{table}

\begin{table}[]
\centering
\setlength{\abovecaptionskip}{0.05cm}
\begin{tabularx}{\textwidth}{|
>{\columncolor[HTML]{ECF4FF}}Y |Y|Y|Y|}
\hline                    & \multicolumn{1}{c|}{\cellcolor[HTML]{ECF4FF}\textbf{PATV ($cm^3$)}} & \cellcolor[HTML]{ECF4FF}\textbf{Deceased} & \multicolumn{1}{c|}{\cellcolor[HTML]{ECF4FF}\textbf{CVD Diagnosis}} \\ \hline
\textbf{Sex}                           & \cellcolor[HTML]{FFFFFF}-50.2*      & \cellcolor[HTML]{9B9B9B}             & -0.30                                                               \\ \hline
\textbf{Age ($years$)}                & 1.70  & 0.03*   & 0.02*                                                               \\ \hline
\textbf{BMI ($kg/m^2$)} & 3.92*                       & \cellcolor[HTML]{9B9B9B}                  & \cellcolor[HTML]{9B9B9B}                                            \\ \hline
\textbf{PATV ($cm^3$)}  & \cellcolor[HTML]{9B9B9B}                 & 0.01*                                     & 0.01*                                        \\ \hline
\end{tabularx}
\caption{Regression coefficients from 3 multivariate regression analyses (n = 391), for the dependent variables listed in each column. \\
Each row shows each regressor considered in the model. (For the dichotomous variable `Sex', females are represented by 1 and males by 0.) Only regressors found to be statistically significant in univariate analyses were considered. 
Entries are the regression coefficients in the specified units, with the asterix (*) denoting significant correlations at $\alpha = 0.01$. PATV: pericardial adipose tissue volume; BMI: body-mass index; CVD: cardiovascular disease.}
\label{tab:results2}
\end{table}

\section{Discussion}
We present PAT-CNN, a CNN able to identify PAT accurately in T2-weighted CMRI across subjects of various sizes and in the presence of different types of pathology. PATV estimated using PAT-CNN's segmentation of CMRI is significantly correlated to the presence of cardiac pathology and is an independent predictor of 1-year mortality, as seen in studies from the literature in which PAT was segmented manually \cite{Fitzgibbons2014EpicardialAssociations}.

\vspace{0.5cm}

\textbf{PAT Segmentation}
PAT is distributed around in the heart in several non-contiguous patches with various shapes, making PAT segmentation more complex than the segmentation of contiguous and less variable structures such as the cardiac chambers themselves. Moreover, when present, pericardial fluid can easily be mistakenly labelled as PAT, as fluid and fat have a similar intensity and appearance on T2-weighted CMR images.

Despite the difficulty of the task, PAT-CNN was able to segment PAT with an average Dice score of 0.74, outperforming U-Net and performing only marginally worse than the average Dice score of 0.76 obtained when comparing the segmentations from different human observers (see Table \ref{tab:results1}).

PAT-CNN uses a 3D Res-UNet previously used to segment heart chambers \cite{Kerfoot2018AutomatedData, Lourenco2021LeftMRI} , with a modified combined Dice-Cross-Entropy loss that outperformed Dice-only and Cross-Entropy-only losses for this task. We therefore show that the 3D Res-UNet has a good performance in CMRI even when segmenting irregular, non-connected regions such as PAT.  

% Even though PAT-CNN can identify the pericardial fat from other parts in CMRI in most cases, errors occurred where there was pleural effusions and heart diseases that involve the accumulation of fluid in images. The reason could be that the water-based fluid has similar pixel values to fat (both are bright given their high T2) in the used T2-weighted MR images. Besides, their positions are usually too close to each other to distinguish. One possible solution is to train the network to segment fat and fluid at the same time, while needs additional computation time. The enlargement of the training dataset could also help in dealing with this problem, as well as avoid overfitting.

\textbf{Statistical Analysis of PATV}
% Females, elderly people and those with higher BMI are more likely to have more PATV. Statistically significant correlations (p-value \textless 0.05) were found between CVD diagnosis / Deceased and PATV, while not with BMI. Thus, PAT is proved to be an independent risk factor for the presence of CVDs and risk of death one-year post-scan. For the next step, more metrics can be investigated. Factors such as subcutaneous fat, adipose tissues’ location, and PFV’s ratio to the heart volume could be meaningful for analysing the risk of CVDs. Moreover, the relationship between pericardial fat and various CVD categories is worth to be explored.

PATV estimated automatically using PAT-CNN shows similar correlations to clinical variables as PATV estimated from CT and echocardiography \cite{Britton2013BodyMortality, Liu2010PericardialStudy}. Females, elderly people and those with higher BMI are more likely to have more PATV. Statistically significant correlations were found between PATV and a diagnosis of cardiovascular disease at the time of scan, as has been reported in other studies \cite{Britton2013BodyMortality}. PATV was also an independent predictor of one-year all-cause mortality in the analysed 319 subjects, as also found on CT \cite{Cheng2010PericardialEvents}. Interestingly, BMI was not a predictor for mortality or the presence of CVD, although it was significantly but weakly correlated with PAT ($\rho_{Pearson} = 0.34$), also as seen in other studies \cite{Britton2013BodyMortality, Henningsson2020QuantificationMRI}.

% Marta added - reread and check is well placed
% \textbf{Limitations}
Our analysis used 1-year all-cause mortality, as we could not obtain information about CVD-related deaths or temporal information about the time to death. Body surface area (BSA) is more commonly used than BMI for the indexing of cardiac function variables. BSA would have been an interesting variable to include in our regression models, but it was not available in our dataset.

\textbf{Future Work}
PAT segmentations, such as those provided by PAT-CNN, provide information beyond that of PATV alone. In the current study, we did not analyse the clinical importance of the location or spatial distribution of PAT, which is likely to have clinical value beyond PAT volume only. We also did not attempt to index PATV by body area or heart size, which may yield biomarkers with further clinical value \cite{Cai2020}. 

Alongside PAT, epicardial adipose tissue (EAT) is also likely to play an important role in cardiovascular pathophysiology and it should be identifiable in CMRI. EAT, however, is expected to deposit in regions of smaller volume and its identification is limited by partial volume effects at MRI's typical spatial resolution. EAT's reliable identification would require MRI at much higher spatial resolution or the use of specialised CMRI sequences such as Dixon techniques \cite{Henningsson2020QuantificationMRI, Kellman2009MultiechoMyocardium}. Dixon technique CMR images are not usually acquired as part of clinical examinations, stressing the importance of our approach to automatically segment PAT from commonly acquired T2-weighted clinical scans.

Pending training and testing across images from different sites and scanners, PAT-CNN can be deployed as a scanner-side analysis tool to automatically estimate PATV. It can also be used to explore the mechanisms of different CVDs by highlighting potential relationships between PAT volume and location and CVD manifestations.

% \textbf{Future Work}
% The study has limitations in several aspects. First, the patient samples are most XXX(?) that limit the generalizability of the subjects. Second, because of the small size of the training dataset, the deep learning model has overfitting issues. Last but not least, we did not take waist circumference or visceral abdominal fat into account; these measures may add incremental information on the effects of local versus systemic adiposity. We also assume that pericardial adipose tissue refers to the adipose tissue deposited around the heart and coronary arteries, while the definition is inconsistent among experts.

\section{Conclusions}
% To conclude, the network PAT-CNN can automatically segment PAT from T2-weighted axial stacks with a Dice score of $0.74 \pm 0.03 mm$. PATV is significantly correlated with the presence of CVD and 1-year mortality, so that can provide information for clinicians.
 
We propose PAT-CNN, a neural network to automatically segment PAT from CMR images. We show that PAT segmentations are reliable, making PAT-CNN ready to be deployed on clinical images to yield biomarkers of potential clinical interest. Using data from 391 patients, we find that PAT volume, estimated using PAT-CNN, is significantly correlated with both a diagnosis of CVD and 1-year all-cause mortality, independently of sex and BMI.

% uncomment after anonymisation can be lifted
\section*{Acknowledgements}
This work was supported by the British Heart Foundation Centre of Research Excellence at Imperial College London (RE/18/4/34215).

\bibliographystyle{splncs04}
\bibliography{references2}

\end{document}